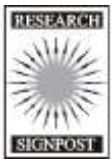

New Developments in Josephson Junctions Research,  Edited by  S. Sergeenkov

# Experimental and theoretical study on 2D ordered and 3D disordered SIS-type arrays of Josephson junctions

Fernando M. Araujo-Moreira[1] and Sergei Sergeenkov[2]

[1]Grupo de Materiais e Dispositivos, Centro Multidisciplinar para o Desenvolvimento de Materiais Cerâmicos, Departamento de Física, Universidade Federal de São Carlos, 13565-905 São Carlos, SP, Brazil
[2]Departamento de Física, CCEN, Universidade Federal da Paraíba, Cidade Universitária, 58051-970 João Pessoa, PB, Brazil

Correspondence/Reprint request: Dr. Fernando M. Araujo-Moreira, Departamento de Física, Universidade Federal de São Carlos, 13565-905 São Carlos, SP, Brazil (E-mail: faraujo@df.ufscar.br )

## Abstract

*By employing mutual-inductance technique and using a high-sensitive home-made bridge, we have thoroughly investigated (both experimentally and theoretically) the temperature and magnetic field dependence of complex AC susceptibility of artificially prepared highly ordered (periodic) two-dimensional Josephson junction arrays (2D-JJA) of both shunted and unshunted Nb–AlO$_x$–Nb tunnel junctions as well as disordered three-dimensional arrays (3D-JJA). This paper reviews some of our latest results regarding the influence of non-uniform critical current density profile on magnetic field behavior of AC susceptibility in 2D-JJA, and the origin of remanent magnetization in disordered 3D-JJAs.*

## 1. Introduction

Many unusual and still not completely understood magnetic properties of Josephson junctions (JJs) and their arrays (JJAs) continue to attract attention of both theoreticians and experimentalists alike (for recent reviews on the subject see, e.g. [1-5] and further references therein). In particular, among the numerous spectacular phenomena recently discussed and observed in JJAs we would like to mention the dynamic temperature reentrance of AC susceptibility [2] (closely related to paramagnetic Meissner effect [3,4]) and avalanche-like magnetic field behavior of magnetization [5,6]. More specifically, using highly sensitive SQUID magnetometer, magnetic field jumps in the magnetization curves associated with the entry and exit of avalanches of tens and hundreds of fluxons were clearly seen in SIS-type arrays [6]. Besides, it was shown that the probability distribution of these processes is in good agreement with the theory of self-organized criticality [7]. It is also worth mentioning the recently observed geometric quantization [8] and flux induced oscillations of heat capacity [9] in artificially prepared JJAs as well as recently predicted flux driven temperature oscillations of thermal expansion coefficient [10] both in JJs and JJAs. At the same time, successful adaptation of the so-called two-coil

Magnetic properties of ordered and disordered Josephson junction arrays      3

mutual-inductance technique to impedance measurements in JJAs provided a high-precision tool for investigation of the numerous magnetoinductance (MI) related effects in Josephson networks [11-14]. To give just a few recent examples, suffice it to mention the MI measurements [12] on periodically repeated Sierpinski gaskets which have clearly demonstrated the appearance of fractal and Euclidean regimes for non-integer values of the frustration parameter, and theoretical predictions [13] regarding a field-dependent correction to the sheet inductance of the proximity JJA with frozen vortex diffusion. Besides, recently [14] AC magnetoimpedance measurements performed on proximity-effect coupled JJA on a dice lattice revealed unconventional behaviour resulting from the interplay between the frustration f created by the applied magnetic field and the particular geometry of the system. While the inverse MI exhibited prominent peaks at $f = 1/3$ and at $f = 1/6$ (and weaker structures at $f = 1/9, 1/12, ..$) reflecting vortex states with a high degree of superconducting phase coherence, the deep minimum at $f = 1/2$ points to a state in which the phase coherence is strongly suppressed.

More recently, it was realized that JJAs can be also used as quantum channels to transfer quantum information between distant sites [15-17] through the implementation of the so-called superconducting qubits which take advantage of both charge and phase degrees of freedom (see, e.g., [18,19] for a review on quantum-state engineering with Josephson-junction devices).

Artificially prepared two-dimensional Josephson junctions arrays (2D-JJA) consist of highly ordered superconducting islands arranged on a symmetrical lattice coupled by Josephson junctions (figure 1), where it is possible to introduce a controlled degree of disorder. In this case, a 2D-JJA can be considered as the limiting case of an extreme inhomogeneous type-II superconductor, allowing its study in samples where the disorder is nearly exactly known. Since 2D-JJA are artificial, they can be very well characterized. Their discrete nature, together with the very well-known physics of the Josephson junctions, allows the numerical simulation of their behavior.

Many authors have used a parallelism between the magnetic properties of 2D-JJA and granular high-temperature superconductors (HTS) to study some controversial features of HTS. It has been shown that granular superconductors can be considered as a collection of superconducting grains embedded in a weakly superconducting - or even normal - matrix. For this reason, granularity is a term specially related to HTS, where magnetic and transport properties of these materials are usually manifested by a two-component response. In this scenario, the first component represents the *intragranular* contribution, associated to the grains exhibiting ordinary superconducting properties, and the second one, which is originated from *intergranular* material, is associated to the weak-link structure, thus, to the Josephson junctions network [20-25]. For single-crystals and other nearly-perfect structures, granularity is a more subtle feature that can be envisaged as the result of a symmetry breaking. Thus, one might have granularity on the nanometric scale, generated by localized defects like impurities, oxygen deficiency, vacancies, atomic substitutions and the genuinely *intrinsic* granularity associated with the layered structure of perovskites. On the micrometric scale, granularity results from the existence of extended defects, such as grain and twin boundaries. From this picture, granularity could have many contributions, each one with a different volume fraction. The small coherence length of HTS implies that any imperfection may contribute to both the weak-link properties and the flux pinning. This leads to many interesting peculiarities and anomalies, many of which have been tentatively explained over the years in terms of the granular character of HTS materials. One of the controversial features of HTS elucidated by studying the magnetic properties of 2D-JJA is the so-called Paramagnetic Meissner Effect (PME), also known as Wohlleben Effect. In this case, one considers first the magnetic response of a granular superconductor submitted to either an AC or DC field of small magnitude. This field should be weak enough to guarantee that the critical current of the intergranular material is not exceeded at low temperatures. After a zero-field cooling (ZFC) process which consists in cooling the sample from above its critical temperature ($T_C$) with no applied magnetic field, the magnetic response to the application of a magnetic field is that of a perfect diamagnet. In this case, the intragranular screening currents prevent the magnetic field from entering the grains, whereas intergranular currents flow across the sample to ensure a null magnetic flux throughout the whole specimen. This temperature dependence of the magnetic response gives rise to the well-known double-plateau behavior of the DC susceptibility and the corresponding double-drop/double-peak of the complex AC magnetic susceptibility [26-31]. On the other hand, by cooling the sample in the presence of a magnetic field, by following a field-cooling (FC) process, the screening currents are restricted to the intragranular contribution (a situation that remains until the temperature reaches a specific value below which the critical current associated to the intragrain component is no longer equal to zero). It has been experimentally confirmed that intergranular currents may contribute to a magnetic behavior that can be either paramagnetic or diamagnetic. Specifically, where the intergranular magnetic behavior is paramagnetic, the resulting magnetic susceptibility shows a striking reentrant behavior. All these possibilities about the signal and magnitude of the magnetic susceptibility have been extensively reported in the literature, involving both LTS and HTS materials [32-35]. The reentrant behavior mentioned before is one of the typical signatures of PME. We have reported its occurrence as a reentrance in the temperature behavior of the AC magnetic susceptibility of 2D-JJA [36,37]. Thus, by studying 2D-JJA, we were able to demonstrate that the appearance of PME is simply related to trapped flux and has nothing to do with manifestation of any sophisticated mechanisms, like the presence of pi-junctions or unconventional pairing symmetry.

The paper is organized as follows. In Section 2 we briefly review the theoretical background for the numerical simulations based on a unit cell containing four Josephson junctions. In Section 3 we describe the influence of non-uniform critical current density profile on magnetic field behavior of AC susceptibility and discuss the obtained results. In Section 4 we study the origin of the so-called remanent magnetization in disordered 3D-JJAs based on both conventional and high-temperature superconductors. And finally, in Section 5 we summarize the main results of the present work.

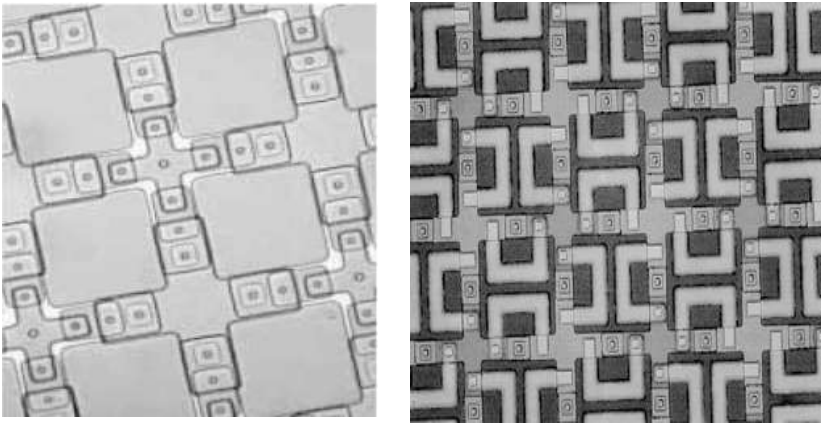

**Figure 1.** Photograph of unshunted (left) and shunted (right) Josephson junction arrays.

## 2. Theoretical background for simulations

We have found that all the experimental results obtained from the magnetic properties of 2D-JJA can be qualitatively explained by analyzing the dynamics of a single unit cell in the array [36,37]. In our numerical simulations, we model a single unit cell as having four identical junctions, each with capacitance $C_J$, quasi-particle resistance $R_J$ and critical current $I_C$. If we apply an external field of the form:

$$H_{ext} = h_{AC}\cos(\omega t) \quad (2.1)$$

then the total magnetic flux, $\Phi_{TOT}$, threading the four-junction superconducting loop is given by:

$$\Phi_{TOT} = \Phi_{EXT} + LI \quad (2.2)$$

where $\Phi_{EXT} = \mu_0 a^2 H_{EXT}$ is the flux related to the applied magnetic field with $\mu_0$ being the vacuum permeability, I is the circulating current in the loop, and L is the inductance of the loop. Therefore the total current is given by:

$$I(t) = I_C(T)\sin\phi_i(t) + \frac{\Phi_0}{2\pi R_j}\frac{d\phi_i}{dt} + \frac{C_j\Phi_0}{2\pi}\frac{d^2\phi_i}{dt^2} \quad (2.3)$$

Here, $\phi_i(t)$ is the superconducting phase difference across the *i*th junction, $\Phi_0$ is the magnetic flux quantum, and $I_C$ is the critical current of each junction. In the case of our model with four junctions, the fluxoid quantization condition, which relates each $\phi_i(t)$ to the external flux, reads:

$$\phi_i = \frac{\pi}{2}n + \frac{\pi}{2}\frac{\Phi_{TOT}}{\Phi_0} \quad (2.4)$$

where *n* is an integer and, by symmetry, we assume that [36,37]:

$$\phi_1 = \phi_2 = \phi_3 = \phi_4 \equiv \phi_i \quad (2.5)$$

In the case of an oscillatory external magnetic field of the form of Eq. (2.1), the magnetization is given by:

$$M = \frac{LI}{\mu_0 a^2} \quad (2.6)$$

It may be expanded as a Fourier series in the form:

$$M(t) = h_{AC}\sum_{n=0}^{\infty}[\chi'_n\cos(n\omega t) + \chi''_n\sin(n\omega t)] \quad (2.7)$$

We calculated $\chi'$ and $\chi''$ through this equation. Both Euler and fourth-order Runge-Kutta integration methods provided the same numerical results. In our model we do not include other effects (such as thermal activation) beyond the above equations. In this case, the temperature-dependent parameter is the critical current of the junctions, given to good approximation by [39,40]:

$$I_C(T) = I_C(0)\left[\frac{\Delta(T)}{\Delta(0)}\right]\tanh\left[\frac{\Delta(T)}{2k_BT}\right] \qquad (2.8)$$

where

$$\Delta(T) = \Delta(0)\tanh\left(2.2\sqrt{\frac{T_C - T}{T}}\right) \qquad (2.9)$$

is the analytical approximation of the BCS gap parameter with $\Delta(0) = 1.76k_BT_C$.

We simulated $\chi_1$ as a function of temperature and applied magnetic fields keeping in mind that $\chi_1$ depends on the geometrical parameter $\beta_L$ (which is proportional to the number of flux quanta that can be screened by the maximum critical current in the junctions), and the dissipation parameter $\beta_C$ (which is proportional to the capacitance of the junction)

$$\beta_L(T) = \frac{2\pi L I_C(T)}{\Phi_0} \qquad (2.10)$$

$$\beta_C(T) = \frac{2\pi I_C C_J R_J^2}{\Phi_0} \qquad (2.11)$$

## 3. Influence of non-uniform critical current density profile on magnetic field response of AC susceptibility in ordered 2D-JJAs

So far, most of the investigations have been done assuming an ideal (uniform) type of array. However, it is quite clear that, depending on the particular technology used for preparation of the array, any real array will inevitably possess some kind of non-uniformity, either global (related to a random distribution of junctions within array) or local (related to inhomogeneous distribution of critical current densities within junctions). For instance, recently a comparative study of the magnetic remanence exhibited by disordered (globally non-uniform) 3D-JJA in response to an excitation with an AC magnetic field $h_{AC}$ was presented [41]. The observed temperature behavior of the remanence curves for arrays fabricated from three different materials (Nb, $YBa_2Cu_3O_7$ and $La_{1.85}Sr_{0.15}CuO_4$) was found to follow the same universal law regardless of the origin of the superconducting electrodes of the junctions which form the array. In this section, through an experimental study of complex AC magnetic susceptibility $\chi(T,h_{ac})$ of the periodic (globally uniform) 2D-JJA of unshunted Nb–AlOx–Nb junctions, we present evidence for existence of the local type non-uniformity in our arrays [42]. Specifically, we found that in the mixed state region $\chi(T,h_{ac})$ can be rather well fitted by a single-plaquette approximation of the over-damped 2D-JJA model assuming a non-uniform (Lorentz-like) distribution of the critical current density within a single junction.

Our samples consisted of $100 \times 150$ unshunted tunnel junctions. The unit cell had square geometry with lattice spacing $a = 46$ μm and a junction area of $5 \times 5$ μm$^2$. The critical current density for the junctions forming the arrays was about 600 A/cm$^2$ at 4.2 K, giving thus $I_C = 150$ μA for each junction. We used the screening method in the reflection configuration to measure the complex AC susceptibility $\chi = \chi' + i\chi''$ of our 2D-JJA (for more details on the experimental technique and set-ups see [36,37]). Figure 2 shows the obtained experimental data for the complex AC susceptibility $\chi(T,h_{ac})$ as a function of $h_{ac}$ for a fixed temperature below $T_C$. As is seen, below 50 mOe (which corresponds to a Meissner-like regime with no regular flux present in the array) the susceptibility, as expected, practically does not depend on the applied magnetic field, while in the mixed state (above 50 mOe) both $\chi'(T,h_{ac})$ and $\chi''(T,h_{ac})$ follow a quasi-exponential field behavior of the single junction Josephson supercurrent (see below).

To understand the observed behavior of the AC susceptibility, in principle one would need to analyze the flux dynamics in our over-damped, unshunted 2D-JJA. However, given a well-defined (globally uniform) periodic structure of the array, to achieve our goal it is sufficient to study just a single unit cell (plaquette) of the array. (It is worth noting that the single-plaquette approximation proved successful in treating the temperature reentrance phenomena of AC susceptibility in ordered 2D-JJA as well as magnetic remanence in disordered 3D-JJA [29,41]. The unit cell is a loop containing four identical Josephson junctions. Since the inductance of each loop is $L = \mu_0 a = 64$ pH and the critical current of each junction is $I_C = 150$ μA, for the mixed-state region (above 50 mOe) we can safely neglect the self-field effects because in this region the inductance related flux [43] $\Phi_L(t) = LI(t)$ is always smaller than the external field induced flux $\Phi_{ext}(t) = B_{ac}(t) \cdot S$. Here I(t) is the total current circulating in a single loop, $S \approx a^2$ is the projected area of a single loop, and $B_{ac}(t) = \mu_0 h_{ac} \cos(\omega t)$ is an applied AC magnetic field. Besides, since the length $\mathcal{L}$ and the width $w$ of each junction in our array is smaller than the Josephson penetration depth $\lambda_j = \sqrt{\dfrac{\Phi_0}{2\pi\mu_0 d j_{c0}}}$ (where $j_{c0}$ is the critical current density of the junction, $\Phi_0$ is the magnetic flux quantum, and $d = 2\lambda_L + \xi$ is the size of the contact area with $\lambda_L(T)$ being the London penetration depth of the junction and $\xi$ an insulator thickness), namely $\mathcal{L} \approx w \approx 5\mu m$ and $\lambda_j \approx 20\mu m$ (using $j_{c0}$ = 600 A/cm$^2$ and $\lambda_L$ = 39 nm for Nb at T = 4.2 K), we can adopt the small junction approximation [43] for the gauge-invariant superconducting phase difference across i*th* junction. Assuming by symmetry that $\phi_1 = \phi_2 = \phi_3 = \phi_4 = \phi_i$, we have:

$$\phi_i(x,t) = \phi_0 + \frac{2\pi B_{ac}(t) d}{\phi_0} \cdot x \qquad (3.1)$$

where $\phi_0$ is the initial phase difference. The net magnetization of the plaquette is $M(t) = SI_S(t)$, where the maximum upper current (corresponding to $\phi_0 = \pi/2$) through an inhomogeneous Josephson contact reads:

$$I_S(t) = \int_0^L dx \int_0^w dy\, j_c(x,y) \cos\phi_i(x,t) \qquad (3.2)$$

For the explicit temperature dependence of the Josephson critical current density we use Eqs.(2.8) and (2.9) from the previous Section.

In general, the values of $\chi'(T, h_{AC})$ and $\chi''(T, h_{AC})$ of the complex harmonic susceptibility are defined via the time dependent magnetization of the plaquette as follows:

$$\chi'(T, h_{ac}) = \frac{1}{\pi h_{AC}} \int_0^{2\pi} d(\omega t) \cos(\omega t) M(t) \tag{3.3}$$

$$\chi''(T, h_{AC}) = \frac{1}{\pi h_{AC}} \int_0^{2\pi} d(\omega t) \sin(\omega t) M(t) \tag{3.4}$$

Using Eqs. (3.1)–(3.4) to simulate the magnetic field behavior of the observed AC susceptibility of the array, we found that the best fit through all the data points and for all temperatures is produced assuming the following non-uniform distribution of the critical current density within a single junction [43]

$$j_c(x, y) = j_{c0}(T) \left( \frac{L^2}{x^2 + L^2} \right) \left( \frac{w^2}{y^2 + w^2} \right) \tag{3.5}$$

It is worthwhile to mention that in view of Eq. (3.2), in the mixed-state region the above distribution leads to approximately exponential field dependence of the maximum supercurrent $I_S(T, h_{AC}) \approx I_S(T, 0) \exp(-h_{AC}/h_0)$ which is often used to describe critical-state behavior in type-II superconductors [27]. Given the temperature dependencies of the London penetration depth $\lambda_L(T)$ and the Josephson critical current density $j_{c0}(T)$, we find that:

$$h_0(T) = \frac{\Phi_0}{2\pi \mu_0 \lambda_j(T) L} \approx h_0(0) \cdot \left( \frac{T_C - T}{T_C} \right)^{1/4} \tag{3.6}$$

for the temperature dependence of the characteristic field near $T_C$. This explains the improvement of our fits (shown by solid lines in figure 2) for high temperatures because with increasing the temperature the total flux distribution within a single junction becomes more regular which in turn validates the use of the small-junction approximation.

       To further study the penetration of the magnetic field in our samples, we have analyzed the so-called susceptibility spectra (see Figs. 3-5). In this case $\chi''$ is a function of $\chi'$, taken at a fixed temperature. The analysis of this spectrum for different temperatures allows follow the evolution of the magnetic flux profile in a particular sample. Ishida and Mazaki [44] have proposed a critical state model based on a superconducting multiconnected structure which has associated a symmetric curve of $\chi''$ as a function of $\chi'$

with its maximum centered at $(-\chi') = 0.5$. On the other hand, Gotoh et al. [45] and Chen et al. [46] have shown that, for other critical states like Bean, exponential, and Kim models, the curve of $\chi''$ as a function of $\chi'$ is asymmetric. The first authors have shown that, for the Bean case, $\chi''(\chi')$ has a maximum value of $\chi'' = 0.239$ at $(-\chi') = 0.375$. Figure 3 is related to an experiment performed at T = 4.2 K. It shows a curve that can be divided into two parts. The first one is symmetric and centered at $(-\chi') = 0.571$, as shown by the solid line. The second one is almost constant and oscillates around $\chi'' \approx 0$. Figure 4 is very similar to Fig. 3; its symmetric part (centered at $(-\chi') = 0.567$) is larger than that in Fig. 3. In both figures the symmetric curve represents the contribution of a multiconnected superconductor, as expected for a 2D-JJA. According to Chen et al. [46], Bean´s model is the low-p limit of the exponential and Kim models. Figure 5, which corresponds to an experiment perfomed at T = 8.0 K, has its best fitting for p = 0.13, so at this temperature both models, Bean and exponential, are equivalents. This equivalence allows us to compare curve (5) to the result of Gotoh et al. [45] deduced for the Bean model, as follows. This figure shows an asymmetric curve with the expected shape for the exponential critical state model [46]. In this case, after its maximum value, $\chi''(\chi')$ should have a linear dependence. has its first part centered at $(-\chi') = 0.262$ and has a maximum value of $\chi'' = 0.239$. As explained before JJA samples never reach complete Meissner at $(-\chi') = 1$, which happens at $(-\chi') = 0.7$ (S.I. units). In this case, the value of $(-\chi') = 0.262$ observed in JJA corresponds to a value of $(-\chi') = 0.374$ associated to superconducting materials in the Bean critical state. Thus, Figure 5 shows the evolution of the critical state from a multiconnected-like to the Bean (or the exponential) critical state model.

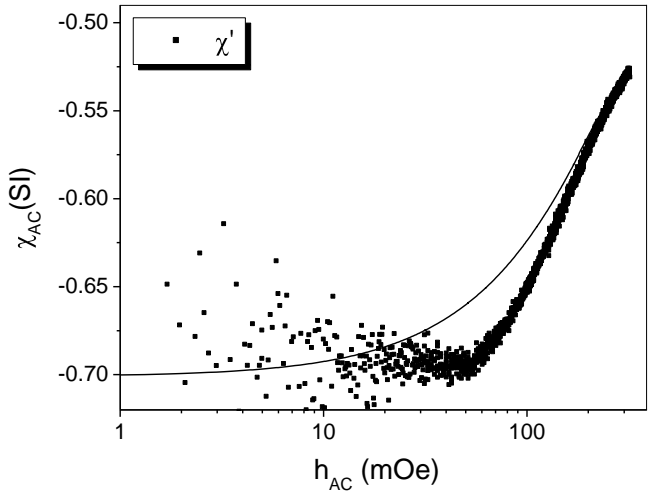

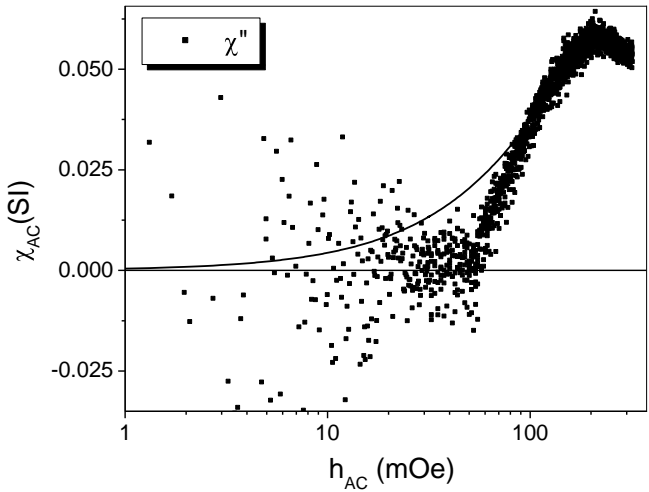

**(a)**

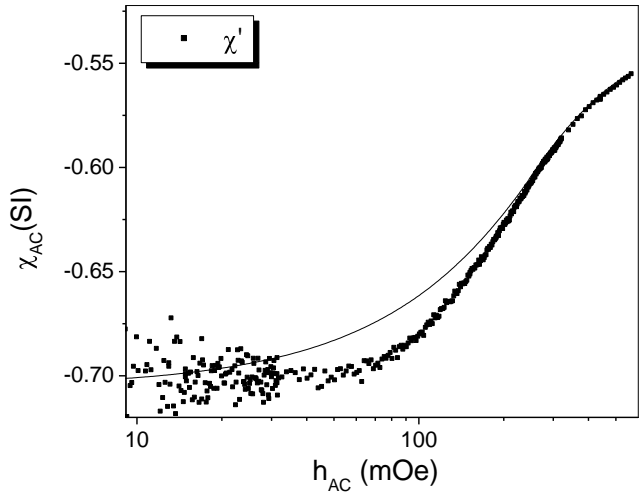

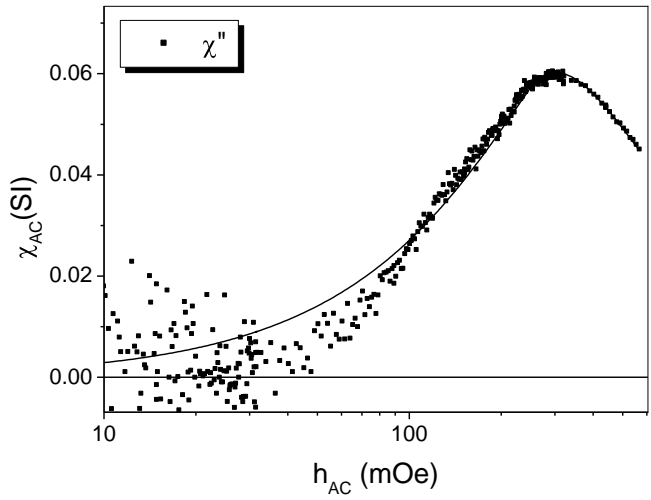

**(b)**

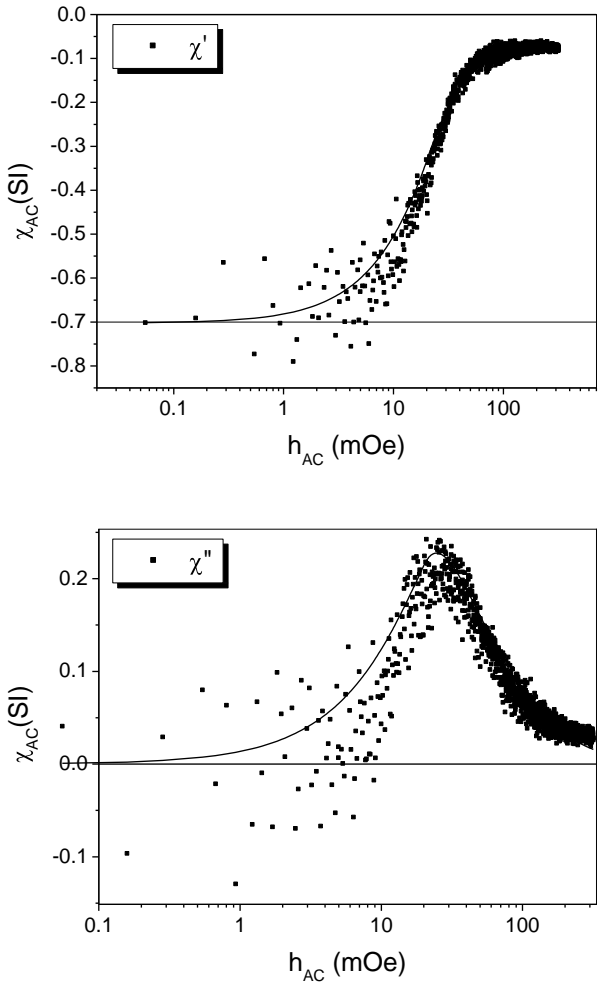

**(c)**

**Figure 2.** The dependence of both components of the complex AC magnetic susceptibilities on AC magnetic field amplitude $h_{AC}$ for different temperatures: (a) T= 4.2 K, (b) T = 6 K, and (c) T = 8 K. Solid lines correspond to the fitting of the 2D-JJA model with non-uniform critical current profile for a single junction (see the text).

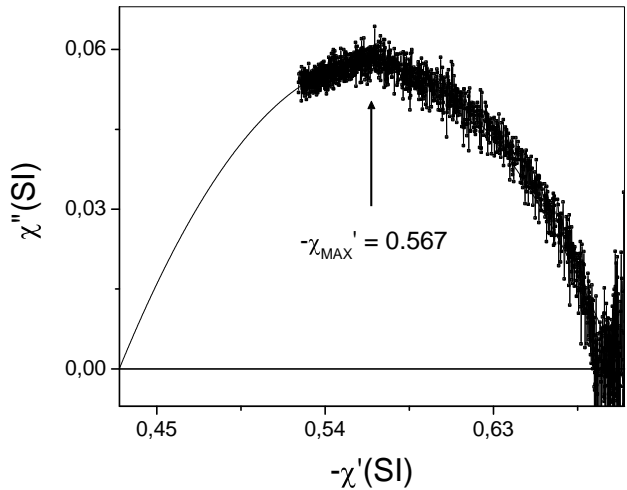

**Figure 3.** Curves for the susceptibility spectra, $\chi''(\chi')$, of an unshunted JJA for T = 4.2 K

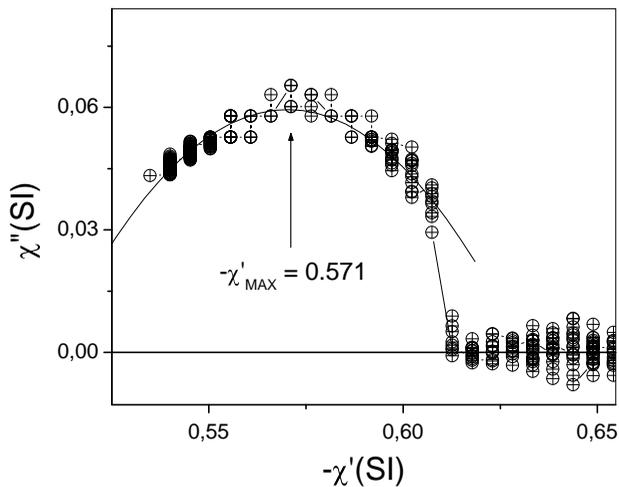

**Figure 4.** Curves for the susceptibility spectra, $\chi''(\chi')$, of an unshunted JJA for T = 6.0 K

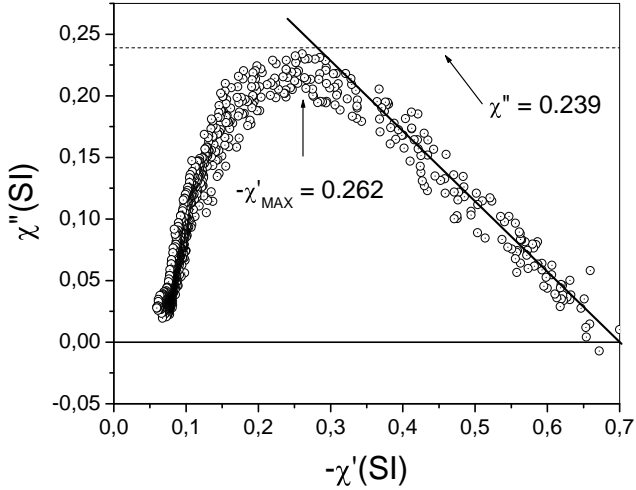

**Figure 5.** Curves for the susceptibility spectra, $\chi''(\chi')$, of an unshunted JJA for T = 8.0 K

## 4. On the origin of the remanent magnetization in disordered 3D-JJAs

The tridimensional disordered Josephson junction arrays (JJAs) fabricated from either conventional (LTS) or high-T_c (HTS) superconductors are known [47,48] to exhibit the so-called temperature-dependent magnetic remanence, $M_R(T)$ upon excitation by a magnetic field. Typically [30], the magnetized state occurs in a rather narrow window of temperatures, the extent of which depends on the critical current, $I_C(T)$, of the junctions. Besides, there is a threshold value for the magnetic field in order to drive the JJA to the state where flux is retained after suppression of the field [30].

In this Section we present a comparative study of three different samples with a rather spectacular remanent behavior and suggest a possible interpretation of the observed temperature dependence of the remanent magnetization of both LTS and HTS tridimensional disordered JJAs. Our analysis shows that all the experimental data can be rather well fitted using the explicit temperature expressions for the activation energy and the inductance-dominated contribution to the magnetization of the array within the so-called phase-slip model [49-51]. Three samples were prepared from selected material, respectively of Nb, $YBa_2Cu_3O_7$ (YBCO) and $La_{1.85}Sr_{0.15}CuO_4$ (LSCO). All three exhibit the predicted remanence and other characteristic features of Josephson arrays. Fabrication routes as well as the experimental routines employed for the magnetic measurements are described elsewhere [47,48]. In short, the corresponding (e.g., niobium) powder was separated according to grain size (using a set of special sieves, with mesh gauges ranging from 38 to 44 μm), then uniaxially pressed in a mold to form a cylindrical pellet of 2.5 mm radius by 2.0 mm height. This pellet is a tridimensional disordered JJA in which the junctions are weakly-coupled grains, i.e., weak-links formed by a sandwich between (Nb) grains and a (Nb-oxide) layer originally present on the grain surface. The measurements were made using a Quantum Design MPMS-5T SQUID magnetometer featuring the regular DC extraction magnetometer and an AC-susceptibility module. The remanence was obtained measuring the sample magnetization after application and removal of a train of sinusoidal pulses. Using the field scan routine we measured the remanent magnetization as a function of the excitation field. For an ordinary superconductor of any kind, from a single crystal to a totally disordered granular sample, the only possibility of a remanence after the application of the AC field would be a residual magnetization due to flux eventually pinned inside the specimen. This contribution, however, is expected to be small and practically independent of the excitation field. We have verified the above characteristics measuring $M_R(h, T)$ for a variety of samples. In particular, the powder used to fabricate our arrays have the typical response of ordinary superconductors, so that the effects described below are entirely due to the formation of the 3D-JJA. The typical experimental data for Nb samples are shown in Figure 6 which clearly demonstrate anisotropic character of the disordered 3D-JJA. The experimental results for all three samples (along with the model fits, see below) are summarized in Figure 7 which suggests that the observed behavior seems to follow a universal temperature pattern, irrespective of the type of superconductor of which the array is made. Let us turn to a possible interpretation of the obtained results. Since the observed remanent magnetization (RM) in our samples (JJAs) appears only below the so-called phase-locking temperature $T_J$ (which marks the establishment of phase coherence between the adjacent grains in the array and always lies below a single grain superconducting temperature $T_C$), it is quite reasonable to assume that origin of RM is related to thermal fluctuations of the phases of the superconducting order parameters across an array of Josephson junctions (the so-called phase-slip mechanism [49-51]). In the present approach we consider the sample as a single plaquette with four Josephson junctions (JJs), each of which is treated via an effective single junction approximation. Within this approximation, the phase-slip scenario yields then

$$\Delta M_R(T) \equiv M(T) - M_R = M_0(T) I_0^{-2}[\gamma(T)/2] - M_R \quad (4.1)$$

for the observed remanent magnetization. Here, $M_0(T)$ is an inductance-induced contribution to the magnetization of the array (see below), $\gamma(T) = U(T)/k_B T$ is the normalized barrier height for thermal phase slippage, $I_0(x)$ is the modified Bessel function, and $M_R = M(T_J)$ is a residual temperature-independent contribution (notice that, according to Eq.(4.1), $\Delta M_R(T_J) = 0$).
For temperatures below $T_J$ (where the main events take place, see Fig.7), the Bessel function can be approximated leading to a simplified version of Eq.(4.1):

$$M(T) = 2\pi M_0(T)[U(T)/k_B T]\exp[-U(T)/k_B T] \quad (4.2)$$

Figure 7 shows the temperature dependence (in reduced units, $\tau = T/T_J$) of the normalized remanent magnetization $m_r(T) = \Delta M_R(T)/\Delta M_R(T_p)$ where $T_p$ is the peak temperature $\Delta M_R(T)$ is defined via Eqs. (4.1) and (4.2). The data for YBCO- and Nb-based JJAs are found to be well fitted with the following explicit expression for the array magnetization:

$$M(t) = A(1-t^4)^{5/2} \exp[-\alpha(1-t^4)] \quad (4.3)$$

where $t = T/T_C$. The best fits through all the data points (shown in Fig.7 by solid and dotted lines for YBCO- and Nb-based JJAs, respectively) using Eq.(4.3) and the known critical parameters:

**YBCO**: $T_C = 90$ K, $T_J = 82$ K, $T_p = 0.88\ T_J$;
**LSCO**: $T_C = 36.5$ K, $T_J = 19.87$ K, $T_p = 0.7\ T_J$;
**Nb**: $T_C = 9.1$ K, $T_J = 8.2$ K, $T_p = 0.92\ T_J$;

yield the following estimates of the model parameters: $\alpha_{YBCO} = 7$, $\alpha_{LSCO} = 2$, and $\alpha_{Nb} = 9$.

Magnetic properties of ordered and disordered Josephson junction arrays 19

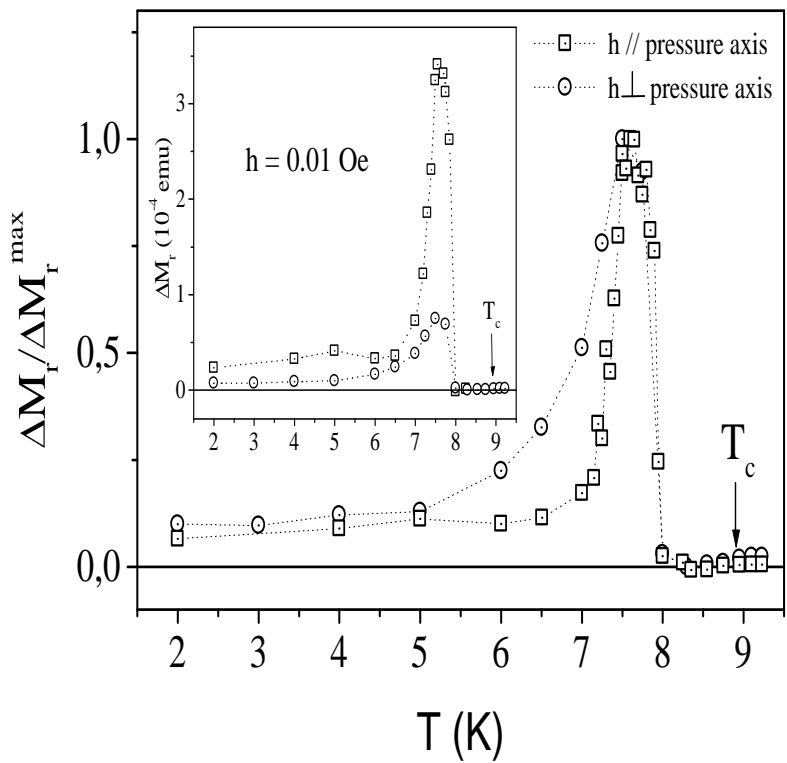

**Figure 6**. Sample anisotropy of a 3D-JJA of Nb, revealed in measurements of the remanence versus temperature for different orientations of the AC excitation field (h). Main graph: data normalized to peak values. Inset: "as measured" data.

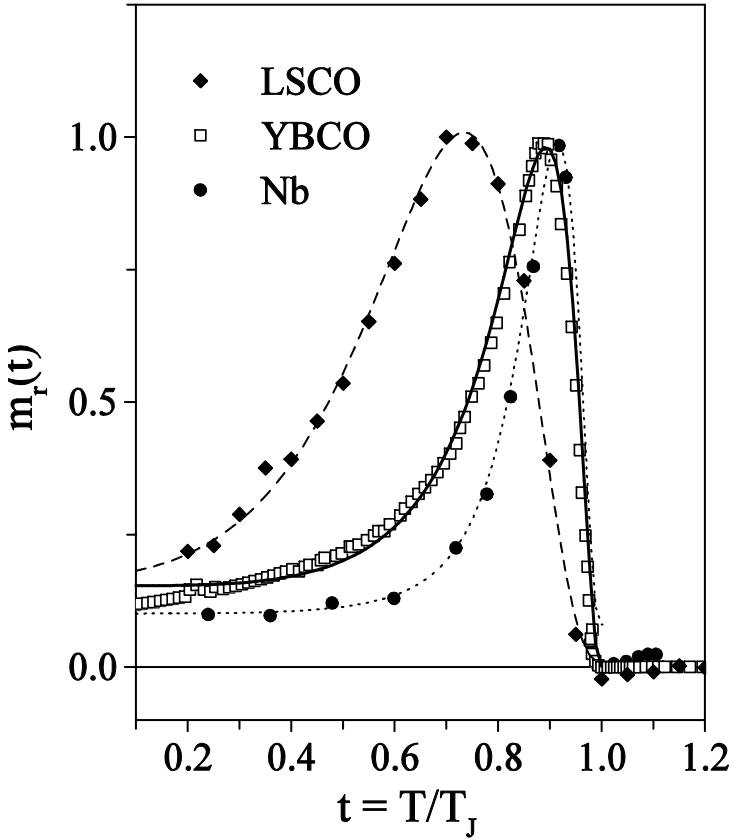

**Figure 7**. Temperature dependence of the normalized remanent magnetization $m_r(T)$, showing the experimental data for three different samples and the corresponding fittings (see text).

To understand the observed behavior of the remanent magnetization, we need to specify the temperature dependencies of the activation energy U(T) and the inductance-dominated contribution $M_0(T)$ to the magnetization of the array. Starting with the YBCO- and Nb-based arrays, it is reasonable to assume that [36,37] $U(T) = \Phi_0 I_C(T)/2\pi$ and $M_0(T)=LI_C(T)/\mu_0 S$, where $I_C(T)$ is an average value of the critical current, L is an average inductance of the Josephson network, S is an effective (in general, temperature-dependent, see below) projected area of the contact, $\Phi_0$ is the flux quantum, and $\mu_0$ is the vacuum permeability. In turn, the temperature dependence of the critical current is dictated by the corresponding dependence of the London penetration depth, namely:

$$I_C(T) = I_C(0)\left[\frac{\lambda_L(0)}{\lambda_L(T)}\right]^2 \tag{4.4}$$

where

$$\lambda_L(T) = \lambda_L(0)\left[1-\left(\frac{T}{T_C}\right)^4\right]^{-1/2} \tag{4.5}$$

Finally, to arrive at the fitting expression given by Eq.(4.3), we have to assume that the projected area S is also temperature dependent (which is not unusual), viz. $S(T) = \pi d(T)l$ with d(T) and l being the thickness and the length of a SIS-type sandwich, respectively ($d(T) = 2\lambda_L(T) + \xi$, where $\lambda_L(T)$ is the London penetration depth and $\xi$ is the thickness of an insulating layer; in ceramics l plays the role of an average grain size $r_g$; typically, $l \gg \lambda_L(T) \gg \xi$).

The above considerations bring about the following relationships between the fitting and the model parametres:

$$A = \left[\frac{LI_C(0)\alpha}{\mu_0 \lambda_L(0)l}\right] \quad \text{with} \quad \alpha = \left[\frac{\Phi_0 I_C(0)}{2\pi k_B T_C}\right] \tag{4.6}$$

At the same time, the phase locking temperature $T_J$, defined via the equation $U(T_J) = k_B T_J$, is related to the critical temperature $T_C$ as follows:

$$T_J = T_C\left(\frac{\alpha}{1+\alpha}\right) \tag{4.7}$$

## 5. Conclusion

In conclusion, in this review paper we presented some of our recent results on novel interesting phenomena related to the magnetic properties of ordered two-dimensional unshunted Nb–AlOx–Nb Josephson junction arrays (2D-JJA) and disordered three-dimensional Josephson junction arrays (3D-JJA) based on conventional (Nb) and high-temperature ($YBa_2Cu_3O_7$ and $La_{1.85}Sr_{0.15}CuO_4$) superconductors. First of all, we demonstrated experimental evidence for the influence of the junction non-uniformity on magnetic field penetration into the periodic 2D array of ordered unshunted Josephson junctions. By using the well-known AC magnetic susceptibility technique, we have shown that in the mixed-state regime the AC field behavior of the artificially prepared array is reasonably well fitted by the single-plaquette approximation of the over-damped model of 2D-JJA assuming inhomogeneous (Lorentz-like) critical current distribution within a single junction. On the other hand, our experimental and theoretical results have demonstrated that the temperature dependence of the magnetic remanence in disordered 3D-JJA is universal, regardless of the origin of the superconducting electrodes of the junctions which form the array.

## Acknowledgment

We thank P. Barbara, C.J. Lobb, A. Sanchez and R.S. Newrock for useful discussions and W. Maluf for his help in running some of the experiments. We gratefully acknowledge financial support from Brazilian Agencies FAPESP and CAPES.